\documentclass{ws-procs9x6}

\begin{document}

\title{$\gamma$-RAYS FROM HEAVY NUCLEI ACCELERATED IN SUPERNOVA REMNANTS}

\author{D. CAPRIOLI$^*$, P. BLASI and E. AMATO}

\address{INAF/Osservatorio Astrofisico di Arcetri,\\
50125 Firenze, Italy\\
$^*$E-mail: caprioli@arcetri.astro.it\\}

\begin{abstract}
We investigate the theoretical and observational implications of the acceleration of protons and heavier nuclei in supernova remnants (SNRs).
By adopting a semi-analytical technique, we study the non-linear interplay among particle acceleration, magnetic field generation and shock dynamics, outlining a self-consistent scenario for the origin of the spectrum of Galactic cosmic rays as produced in this class of sources. 
Moreover, the inferred chemical abundances suggest nuclei heavier than Hydrogen to be relevant not only in the shock dynamics but also in the calculation of the $\gamma$-ray emission from SNRs due to the decay of neutral pions produced in nuclear interactions. 
\end{abstract}

\keywords{cosmic rays, acceleration of particles, SNR, $\gamma$-rays}

\bodymatter

\section{Cosmic rays and supernova remnants}
For more than 70 years scientists have been regarding supernova remnants (SNRs) as the most plausible sources for Galactic cosmic rays (GCRs), but only in recent times the theoretical comprehension of the ongoing processes have made important steps forward (see also P. Blasi's contribution in this volume). 
Once a model for CR propagation in the Milky Way is assumed, it is possible to infer from observations at Earth the spectra of single chemical species expected at the sources.
In this work we investigate a scenario in which Galactic SNRs are responsible for the acceleration of protons and heavier nuclei (hereafter HN), trying to disentangle the solid, physical ingredients and the phenomenological recipes which have to be included in order to account for many observational constraints.
In particular we show that, according to this \emph{SNR paradigm} for the origin of GCRs, HN may play a fundamental role in the shock dynamics and also contribute in a non-negligible way to the $\gamma$-ray emission from SNRs.

We consider here a semi-analytical approach to the problem of non-linear diffusive acceleration of particles at shocks (NLDSA) following the basic implementation put forward in Ref.~\refcite{ab05, ab06, feb}, also including the effects of the magnetic field amplification on the shock dynamics \cite{jumpkin} and the pressure of the most abundant CR species\cite{nuclei}.
The evolution of a remnant in a homogeneous circumstellar medium is followed in a quasi-stationary way as in Ref.~\refcite{tmk99} and coupled with the acceleration of particles at the forward shock as in Refs.~\refcite{nuclei,crspec}.
For the details of the computational apparatus, the reader may refer to the papers above. 
Here, we only would like to highlight how such a semi-analytical approach to NLDSA is typically much faster than, and as much rigorous as, other fully numerical or Monte Carlo methods for non-relativistic shocks\cite{comparison}, and therefore it is very useful for including multi-specie CRs or, in general, for studying problems with a wide range of environmental parameters.

In our calculations protons are injected into the acceleration process from the thermal bath through a ``thermal-leakage'' mechanism\cite{bgv05}, which is sensitive to the shock dynamics and tends to suppress the number of injected particles when the the acceleration becomes more and more efficient. 
The amounts of injected HN are tuned by hand relatively to protons to reproduce the abundances measured at Earth.
This recipe, while including a reasonable feedback which self-regulates the efficiency of the injection, has to be regarded as a necessary phenomenological description since, unfortunately, the HN injection at SNR shocks is still poorly understood for two main reasons.
First, injection strongly depends on the charge/mass ratio and it is very difficult to follow the degree of ionization of heavy atoms during their acceleration; second, refractory elements (Mg, Si, Fe,...) are thought to be injected as a result of sputtering of accelerating dust grains\cite{dust1,dust2}.
The latter phenomenon, though very hard to deal with quantitatively, is however expected to produce largely suprathermal (but still non-relativistic) ions able to cross the shock from downstream to upstream because of their large gyroradii. For the same reason, partially ionized heavy atoms are expected to be preferentially injected, in agreement with the relative abundances measured in GCRs\cite{dust1,dust2}.
Nevertheless, since our abundances are tuned on the relativistic region of the GCR spectrum, we can bypass the problem above and make safe predictions about the role of HN in the shock dynamics and in the SNR $\gamma$-ray emission.

At any given time the shock dynamics is regulated by the non-linear interplay between particle acceleration, occurring via first-order Fermi mechanism, and magnetic field amplification, which we model as due to resonant streaming instability excited by all the accelerated particles\cite{bell78}.
On one hand, the pressure in CRs diffusing around the shock leads to the formation of a precursor which slows down the incoming fluid and tends to make the shock weaker while, on the other hand, the pressure in the shape of self-generated magnetic turbulence may become comparable to, or even larger than, the gas pressure upstream, preventing an excessive modification of the velocity profile.
This {\it magnetic feedback}\cite{jumpl} has proved itself to be a very common mechanism able to account, at the same time, for both the level of magnetization and the hydrodynamics inferred by multi-wavelength studies of young SNRs \cite{jumpkin}. 
In addition, when the velocity of the scattering centres, which we assume to be transverse Alfv\'en waves as predicted by the quasi-linear theory of resonant streaming instability, becomes a non-negligible fraction of the fluid velocity, the compression ratios actually felt by the fluid and by the accelerated particles are no longer the same. 
A physically motivated account for the relative velocity between the fluid and the waves leads to conclude that the more efficient the magnetic field amplification (the larger the Alfv\'en velocity) is, the steeper the spectra of the accelerated particles are (see e.g. \S7 of Ref.~\refcite{nuclei}). 

\section{Particle escape from SNRs}
In order to explain the observations at Earth, one has to account for \emph{when} and \emph{how} accelerated particles leave the source and become CRs. 
There are three main mechanisms playing a potential role in the problem: escape from upstream, escape from downstream and release at the ``death'' of the remnant (after having undergone adiabatic losses).
In general, the total CR flux provided by a single source is given by the convolution over its life of these three contributions. Let us discuss them in some more detail.

During the ejecta-dominated stage the shock velocity $V_{sh}$ is roughly constant: the magnetic field is expected to increase, in turn leading to more and more energetic particles. 
In such a situation no particle can diffuse away from upstream.
During the Sedov-Taylor stage, instead, both the shock velocity and the magnetization level decrease: the diffusion length increases as $1/V_{sh}\propto t^{3/5}$ and hence the highest-energy particles are no longer able to make it back to the shock, since it moves as $R_{sh}\propto t^{2/5}$\cite{escape}. 
Quantitatively speaking, we assume that at any moment the system can confine only particles with diffusion length $D(E,\delta B)/V_{sh}\leq \chi R_{sh}$, where $D(E,\delta B)$ is the Bohm diffusion coefficient in the amplified field $\delta B$, by imposing the distribution function of the accelerated particles to vanish for radii larger than $(1+\chi) R_{sh}$ (see also Ref.~\refcite{feb}).

Clearly, most of accelerated particles are advected behind the shock and, trapped in the expanding shell, cannot do but losing energy adiabatically. 
If, for any reason (like, for instance, the presence of strong density or magnetic patterns in the circumstellar medium), the SNR shell were ``broken'', preferential conduits might channel particles from downstream into the Galaxy.
Such a contribution is in principle very hard to model, thus we assume that, at any given time, a 10\% of the downstream accelerated particles is injected into the Galaxy, contributing to the diffuse GCRs\cite{crspec}.
Finally, the remainder of the advected particles should be released at the ``death'' of the SNRs, which is also quite difficult to define: it might coincide with the beginning of the radiative stage, but actually it should be more closely related to the damping of the magnetic field and hence to the transition from a SNR-like to a Galactic-like regime for the particle diffusion.

We consider here a ``benchmark'' SN explosion releasing 1.4 solar masses with total kinetic energy $10^{51}$erg in the homogeneous circumstellar environment, taken with temperature $10^{6}{\rm K}$ and particle density $\rho_{0}=0.01m_{\rm H}$cm$^{-3}$. 
The remnant evolution is thus followed as in Ref.~\refcite{tmk99}, Tab.~7, and the SNR is imposed to die at the end of the Sedov stage. 
The transport in the Galaxy is then calculated within a simple {\em leaky-box} model, assuming a halo height of 3.5 kpc, a Galaxy radius of 10 kpc and a residence time $\tau_{esc}(R)=20 (R/10 {\rm GV})^{-0.55}$ Myr, where $R$ is the nucleus rigidity.
When spallation (mainly relevant at low energies and for HN) is also included\cite{hoer+07}, the spectra of the GCRs is recovered in terms of slope, normalization and cut-off (see Figs.~3,4 in Ref.~\refcite{nuclei}). 

It is however worth keeping in mind that the recipe linking magnetic field amplification and the velocity of the scattering centers is rather phenomenological and it has not been thoroughly calculated from first principles, yet.
Remarkably, at the moment it represents the only proposed effect able to produce source spectra steeper than $E^{-2}$ (see also Ref.~\refcite{pzs10}) and therefore consistent with the inferred value of $\delta=0.3-0.6$ for the scaling of the Galactic residence time $\tau_{esc}\propto E^{-\delta}$.
In fact, in Ref.~\refcite{berevolk} such an effect is not included and $\delta=0.75$ is required to recover the GCR spectrum. 

\section{The role of heavy nuclei}
The chemical abundances fitting the GCR data at Earth are self-consistently put in the computational apparatus in a recursive way until consistency is reached, since the non-linear nature of the acceleration process does not allow an {\em a priori} exact determination of the proton/HN ratios.  
Provided that the relative abundances do not depend on time, such a process allow us to investigate the role of HN at any given time of the SNR evolution. 

\begin{figure} 
\psfig{file=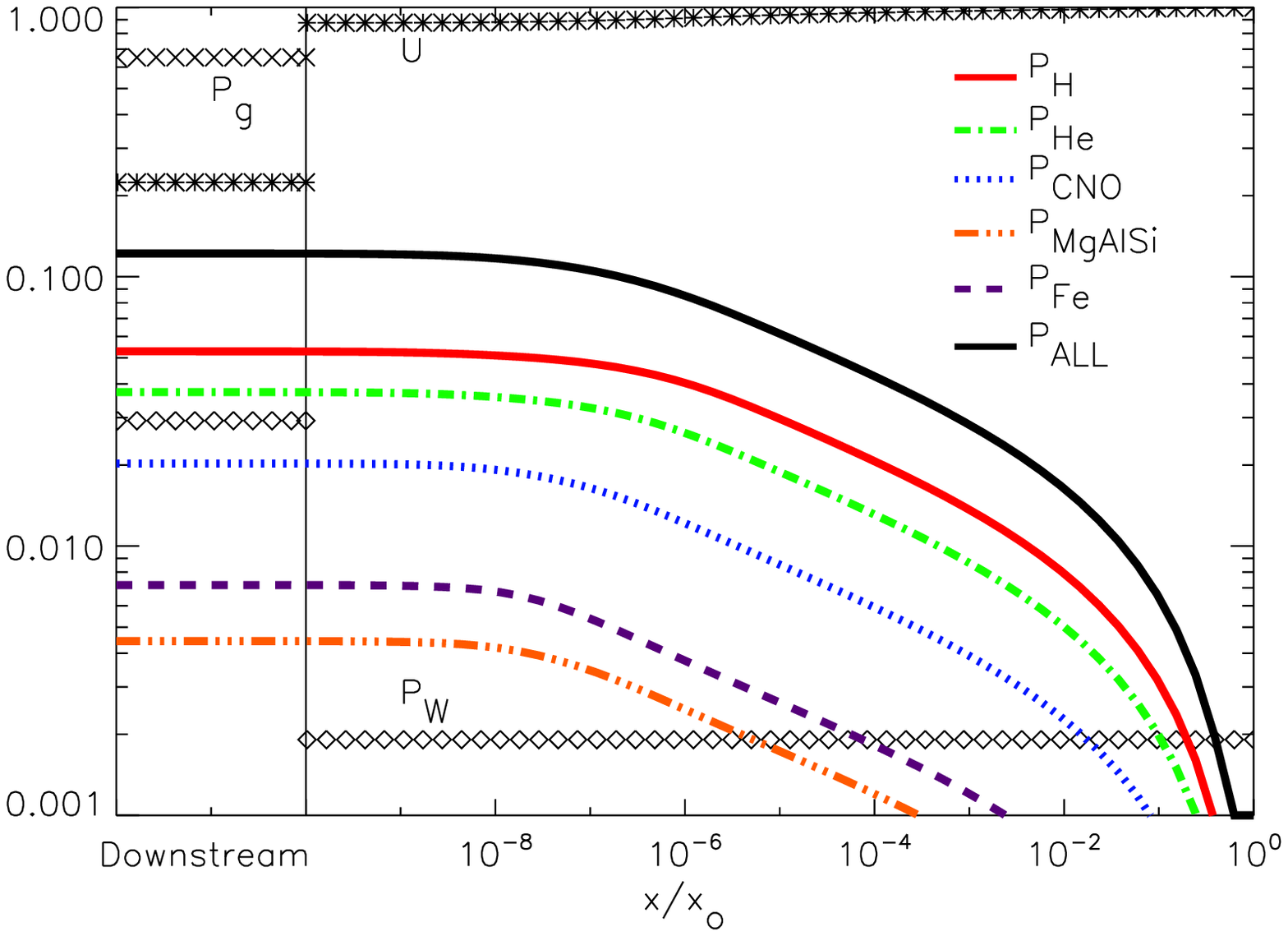,width=2.2in} 
\psfig{file=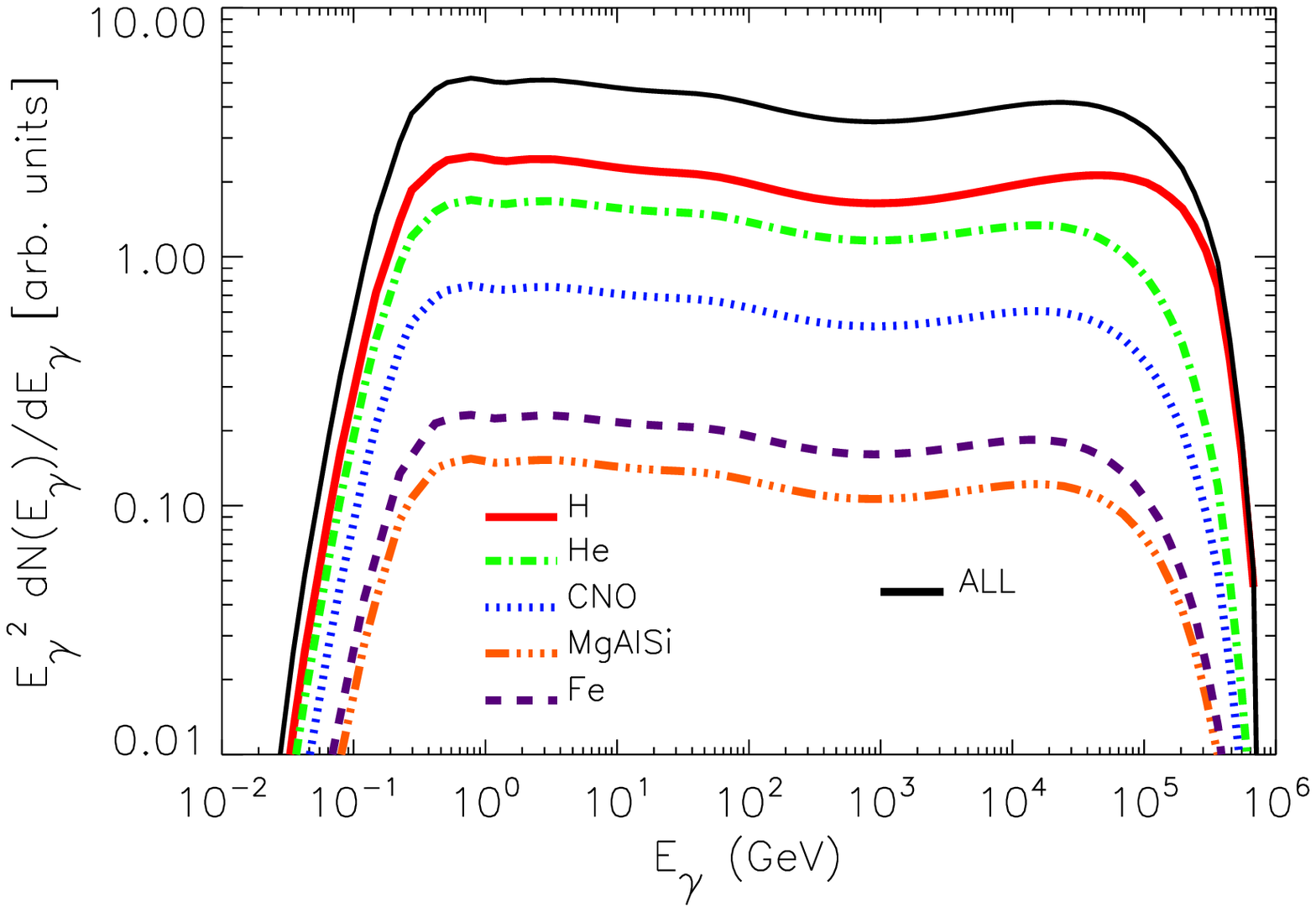,width=2.2in}
\caption{Left panel: spatial profile of the modified shock. See text for the details of the different curves.
Right panel: $\gamma$-ray spectrum from $\pi^{0}$ decay produced in nuclear collisions, in arbitrary units. Both panels correspond to a SNR age of about 2000 yr.} 
\label{fig} 
\end{figure}

In the left panel of Fig.~\ref{fig} a snapshot of the shock profile at the age $t=2000$ yr (beginning of the Sedov phase) is shown, in units of $x_{0}=\chi R_{sh}\simeq 0.2 R_{sh}$, i.e. the distance of the free-escape boundary from the shock. 
$U(x)$ is the local velocity in units of $V_{sh}$, $P_{g}(P_{w})$ is the gas (magnetic field) pressure and the other curves correspond to the contributions to the pressure by accelerated particles of different species, as in the legend; all pressures are normalized to $\rho_{0}V_{sh}^{2}$.

It is easy to see that more than 10\% of the bulk pressure is channelled into accelerated particles, and that HN contribute as much as the protons to the shock dynamics. 
Most of their contribution comes from He nuclei, but heavier elements still account for about 25\% of $P_{\rm ALL}$.
As outlined above, the pressure in magnetic turbulence $P_{w}$ turns out to be larger than the gas pressure $P_{g}<10^{-4}$, and thus the precursor is dynamically dominated by the interplay between CRs and magnetic field\cite{jumpl,jumpkin}.
Moreover, since $P_{w}\propto P_{\rm ALL}$, we can postulate that HN play a non-negligible role also in the amplification of the background magnetic field via streaming instability.

Another interesting consequence of the acceleration of HN is that they may make an impact on the hadronic $\gamma$-ray emission from SNRs, i.e. the emission due to the decay of neutral pions produced in nuclear collisions between relativistic particles and background gas\cite{dav94}.
For distributions of accelerated particles $\propto E^{-2}$, the contribution to the $\gamma$-ray flux by different chemicals is simply proportional to their relative abundances (see also \S5 in Ref.~\refcite{nuclei}). This means that, for the chemical composition determined above, HN contribute in a substantial way to the $\gamma$-ray flux from a SNR (more than a factor 2 for the case depicted in Fig.~\ref{fig}). 
The maximum $\gamma$-ray energy produced by all the HN is typically half the one produced by protons, since the acceleration is rigidity-dependent while the energy of the secondaries scales with the energy per nucleon.
This effect leads to an alteration in the shape of the cut-off, a signature probably too hard to be discriminated by observations (ALL vs H curve in the right panel of Fig.~\ref{fig}).

Very generally, the inclusion of HN implies that the observed $\gamma$-ray flux from a given SNR may be achieved with, say, half the density of the target gas with respect to a case accounting for accelerated protons only.
As a consequence, the HN contribution is of great importance when the density of the circumstellar medium is deduced by the level of $\gamma$-rays of hadronic origin. 
A different estimate of the circumstellar density, in principle, leads also to a different modelling of the SNR evolution, affecting the inferred age, radius, expansion rate and, eventually, distance of the object.
Even more interestingly, a reduced circumstellar density would also imply a strong suppression ($\propto \rho^{2}$) of the expected thermal emission of the shocked plasma, both in terms of continuum and lines. 
Possible HN overabundances, with respect to the ``standard'' values above, might as well account for the lack of detection of thermal emission in single $\gamma$-ray-bright SNRs as, for instance, RX J1713.7-3946.

\end{document}